\pdfoutput=1
\documentclass[3p,a4paper,numbers,sort&compress,%
review%
]{elsarticle}%
\journal{Journal of Aerosol Science}
\usepackage{amsfonts,amsmath,amssymb}
\usepackage[T1]{fontenc}
\usepackage{graphicx}
\usepackage[utf8]{inputenc}
\usepackage{microtype}
\usepackage[textsize=footnotesize,textwidth=18mm]{todonotes}
\usepackage{xcolor}
\usepackage{hyperref}
\usepackage[todos]{CMImacros}

\renewcommand{\topfraction}{0.99}

\begin{document}
\renewcommand{\topfraction}{0.99}
\begin{frontmatter}
   \title{Optimizing aerodynamic lenses for single-particle imaging}%
   \author[cfel,uhhphys]{Nils Roth}%
   \author[cfel,uhhcui]{Salah Awel}%
   \author[cfel,uhhcui]{Daniel A.\ Horke}%
   \author[cfel,uhhphys,uhhcui]{Jochen Küpper}\corref{correspond}%
   \cortext[corresponding]{Author to whom correspondence shall be addressed:}%
   \ead{jochen.kuepper@cfel.de}%
   \ead[url]{https://www.controlled-molecule-imaging.org}%
   \address[cfel]{Center for Free-Electron Laser Science, Deutsches Elektronen-Synchrotron DESY,
      Notkestrasse 85, 22607 Hamburg, Germany}%
   \address[uhhphys]{Department of Physics, Universit\"at Hamburg, Luruper Chaussee 149, 22761
      Hamburg, Germany}%
   \address[uhhcui]{The Hamburg Center for Ultrafast Imaging, Universit\"at Hamburg, Luruper
      Chaussee 149, 22761 Hamburg, Germany}%
\begin{abstract}
   A numerical simulation infrastructure capable of calculating the flow of gas and the trajectories
   of particles through an aerodynamic lens injector is presented. The simulations increase the
   fundamental understanding and predict optimized injection geometries and parameters. Our
   simulation results were compared to previous reports and also validated against experimental data
   for 500~nm polystyrene spheres from an aerosol-beam-characterization setup. The
   simulations yielded a detailed understanding of the radial phase-space distribution and
   highlighted weaknesses of current aerosol injectors for single-particle diffractive imaging. With
   the aid of these simulations we developed new experimental implementations to overcome current
   limitations.
\end{abstract}
\end{frontmatter}

\section{Introduction}
\label{sec:introduction}
Single-particle diffractive imaging (SPI) is one of the key applications enabled by the advent of
x-ray free-electron lasers (XFELs)~\cite{Bogan:NanoLett8:310, Seibert:Nature470:78}. Short-duration
XFEL pulses were predicted to allow the collection of diffraction patterns from radiation-sensitive
samples without resolution limitations due to radiation damage~\cite{Neutze:Nature406:752,
   Nass:JSR22:225}, although some open questions remain~\cite{Lorenz:PRE86:051911,
   Ziaja:NJP14:115015, Nass:JSR22:225}. A series of two-dimensional diffraction patterns of randomly
oriented isolated particles can be used to reconstruct the full three-dimensional structure, without
the need for large highly ordered crystalline samples~\cite{Bogan:AST44:i, Seibert:Nature470:78,
   Ekeberg:PRL114:098102}.

As every intercepted particle is destroyed by the intense x-ray pulse~\cite{Chapman:NatMater8:299},
a new and preferably identical sample particle has to be delivered into every pulse. This can be
achieved with aerosolized particle beams, which, furthermore, offer significantly reduced background
levels compared to liquid jet based delivery methods~\cite{DePonte:JPD41:195505, Awel:JACR51:133}.
The most widespread aerosol injectors for SPI experiments are aerodynamic lens stacks
(ALS)~\cite{Bogan:NanoLett8:310, Hantke:NatPhoton8:943}. However, other aerosol injectors, \eg,
convergent nozzles, have also been demonstrated~\cite{Kirian:SD2:041717, Awel:OE24:6507,
   Awel:JACR51:133}. One of the limiting factors for SPI is the collection of a sufficient number
of strong diffraction patterns~\cite{Bergh:QRB41:181, Fung:Nature532:471}. Overcoming this
limitation requires the delivery of high-density particle streams in order to maximize the number of
x-ray pulses intersecting a particle and producing a measurable diffraction pattern. Using current
aerosol injectors, hit fractions, \ie, the fraction of x-ray pulses that hit at least one particle,
up to $79~\%$ could be achieved~\cite{Hantke:NatPhoton8:943}. However, this contains pulses
interacting with multiple particles, faint hits far from the x-ray maximum intensity, as well as
hits from background particles. This leads to hit rates, \ie, usable diffraction patterns containing
a bright image from a single isolated target particle, of below $5~\%$ and, hence, long measurement
times and excessive sample consumption. Furthermore, these current studies have been undertaken with
x-ray-focal-spot sizes on the order of a few micrometers. This comparatively soft focusing of the
XFEL pulse does not yield the photon intensity required for measurable SPI diffraction signal to
high scattering angles or from small samples~\cite{Barty:ARPC64:415}. This requires nanofocused
x-ray beams with focal spot sizes on the order of 100~nm, where the hit-rate achievable with current
aerosol injectors is typically below $0.05~\%$.

\begin{figure}
   \centering
   \includegraphics{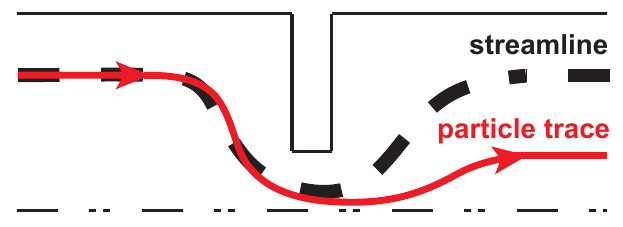}
   \caption{Schematic of particles being contracted toward the centerline by a gas stream through
      an orifice. The orifice is cylindrical symmetric around the dashed line. The trajectories of
      particles past the orifice vary from the streamlines due to particle inertia, leading to an
      aerodynamic lensing effect.}
   \label{fig:adl}
\end{figure}

As particles distribute stochastically in the aerosol beam, the probability for them to be within
the x-ray interaction volume depends on the local particle density, necessitating highly collimated
or focused particle streams. Robinson predicted in 1956 that in real, irrotational and
incompressible gas flow past an obstacle, the density of small particles within the flow can
increase while passing the obstacle~\cite{Robinson:CPAM9:69}. This mathematical description was
later extended and used, supported by numerical simulations, to describe particles flowing in a tube
through an orifice~\cite{Liu:AST22:293}. Under the right conditions, in what is now known as an
aerodynamic lens (ADL), the particles concentrate at the center of the tube, as illustrated in
\autoref{fig:adl}.

A detailed numerical characterization of an individual ADL was presented in
2002~\cite{Zhang:AST36:617}, which was later extended to an entire ALS~\cite{Zhang:AST38:619}.
Numerical simulations for an ALS to focus particles with diameters below 30~nm~\cite{Wang:AST39:624}
led to a simple design tool that predicts the required lens dimensions to focus a specified range of
particle sizes at given flow conditions~\cite{Wang:AST40:320}. Based on this, further numerical
simulations have adapted ALS to specific needs~\cite{Lee:JAS39:287, Meinen:AST44:316}. Although ALS
have been used for, \eg, ultrafast electron imaging experiments on
nanoparticles~\cite{Zherebtsov:NatPhys7:656}, they are predominantly used in aerosol mass
spectrometry~\cite{Canagaratna:MSR26:185}. Here, the main goal is to contract a large range of
particle sizes and ensure a high transmission. The ``Uppsala'' ALS, a widely used standard injector
for SPI~\cite{Hantke:NatPhoton8:943}, was designed based on the same principle. Therefore, it can
deliver collimated particle beams for a large range of particle sizes ($0.1$--$3$~\um) without
changing the apparatus. However, the possibility to optimize for a specific particle size is
limited, and the pressure before the ALS is the only tunable parameter.

These design principles differ from the requirements of SPI experiments, where highly collimated
beams of only one particular particle size are needed, and even desirable in order to increase
sample purity. Additionally, the final particle beam diameter should be matched to the x-ray focal
size. To enable the transmission of a wide range of particle sizes requires the use of several
orifices within the ALS. This increases the complexity of the setup and the individual orifices are
designed for different particle sizes, making some of them counterproductive for producing a
high-density beam of a well-defined particle size.

Here, we present a detailed numerical simulation environment to understand and to quantitatively
model the underlying fundamental processes occurring within an ALS and to further optimize these
systems to meet the requirements of SPI experiments. In particular, we aim to design an ALS
optimized for focusing a single particle size to the smallest possible beam diameter, while keeping
the experimental setup simple and easily adaptable for different samples.

\section{Methods}
\label{sec:methods}
\subsection{Numerical Simulation}
\label{sec:simulation}
Optimizing the geometry of an ADL requires investigating a large parameter space, such as dimensions
of orifices and transport tubes, making experimental characterization and optimization impractical.
Instead, we implemented numerical simulations to predict the behavior of particles within the ADL.
Furthermore, these simulations allow the extraction of phase-space distributions of particles at any
position within the device. Within all simulations we assumed that, (i) particles in the flow field
have no influence on the flow field itself and (ii) particles do not interact with each other. This
implies that the flow field and particle trajectories can be calculated separately, and that each
particle can be simulated independently. These assumptions significantly reduce computational cost,
and are easily justified considering the typical pressures in an ALS. The helium (or carrier gas)
pressure is $\sim$1~mbar (number density of $\sim10^{16}$~atoms/cm$^3$), while the density of
particles usually does not exceed $10^{10}$~particles/cm$^3$.

For an accurate description of the ALS and produced particle beams, the phase-space distribution of
particles at the inlet of the injector is a crucial parameter. This distribution is typically
defined by either the aerosol source or, more commonly, by a differential pumping stage used to
reduce the gas-load and to control the pressure upstream of the ALS. A common arrangement for such a
pumping stage is a set of two skimmers, oriented with the tips facing each other, as described in
detail in \autoref{sec:exp_setup} and shown schematically in \autoref{fig:uppsalalens}. Since there
is no experimental data available for the phase-space distribution of particles before they enter
the ALS, the initial particle conditions are evaluated through simulations of the flow through the
skimmers.

\subsubsection{Flow field}
We simulated the flow field of the carrier gas using a finite-element
solver~\cite{Comsol:Multiphysics:5.3} to solve the Navier-Stokes equations. The geometry and flow
were assumed to be axisymmetric about the central axis, and the flow solved in two dimensions
($r,z$). The flow was treated compressible and viscous, and the calculation iterated until converged
to a steady-state solution. Additional properties of the flow field are indicated by three
dimensionless quantities, the Reynolds number \Reyn, the Knudsen number \Knud, and the Mach number
\Mach. \Reyn is defined as the ratio of inertial to viscous forces. Typically, Reynolds numbers in
an ALS are below 10, indicating that no instabilities are present in the flow field, which we thus
solve assuming laminar flow. \Knud is defined as the ratio of the mean free path to a characteristic
length. Inside the ALS the pressure is usually on the order of 1~mbar, corresponding to a mean free
path around 70~\um for helium at room temperature. Compared to the dimensions of apertures (a few
mm) this results in $\Knud<0.01$ and there are enough collisions with the background gas to treat
the flow as continuum and the Navier-Stokes equations hold. Upstream of the ALS, before
   and throughout the two skimmer setup, the pressure is even higher, hence $\Knud\ll0.01$. In the
vacuum chamber, on the other hand, at helium pressures of $10^{-2}$~mbar or less, the mean free path
is around 10~mm, hence $\Knud>0.01$, and the continuum flow model breaks down. With such a large
mean free path particles rarely collide with the background gas, such that there is no momentum
exchange with the flow field anymore. In between these regimes there is a transition region which is
difficult to model. However, this transition between regimes occurs rapidly after particles exit the
tip of the injector, and here we assumed a sudden stop of the continuum flow regime and an immediate
change to the molecular flow regime. In the latter, particles were assumed to propagate
collision-less and with constant velocity. The ratio of the velocity of the flow to the speed of
sound in the fluid is the Mach number, \Mach. For $\Mach>0.3$ effects due to the
   compressibility of the fluid start to occur, such as pressure waves and cooling of the fluid.
   Here, such high \Mach are reached between the two skimmers and downstream of the injector tip.
   Pressure waves are still properly described by the Navier-Stokes equations. However, our
   numerical simulation approach, finite element methods, necessitates the use of stabilization
   methods, which add artificial diffusion in order to avoid numerical instabilities, \eg,
   oscillations in the solution. While these might wash out the position and velocity of the
   potentially occurring pressure waves, the position of the second skimmer is closer to the
   first skimmer exit than the calculated distance of the Mach disc, where these pressure waves
   are supposed to be located. In Addition, downstream of
   the injector tip and between the two skimmers particles are fast and have a high intertia. Hence,
   the effect of the spiky features caused by high \Mach are assumed to have a limited influence on
   the overall particle trajectories, especially at the injector tip, where the continuum flow
   breaks. Nevertheless an accurate treatment of the flow including thermodynamic coupling might be
   able to further improve the simulation results.

\subsubsection{Particle traces}
\label{sec:particle-traces}
Particle trajectories were simulated, within a given steady-state flow field, with a homebuilt
\texttt{python} code that uses a real-valued variable-coefficient ordinary-differential-equation
solver. The code interpolates the given pressure and velocity fields and calculates the forces,
described by Stokes' law, acting on a particle of given size at each time step.
Additional corrections can arise depending on the particle's Reynolds ($\Reyn_\text{p}$) and Knudsen
($\Knud_\text{p}$) numbers. These are defined identical to the fluid case, but with the
characteristic length given by the particle diameter. $\Reyn_\text{p}$ is very small inside the lens
($\Reyn_\text{p}<1$) and can be neglected. $\Knud_\text{p}$, however, cannot be ignored as at low
gas densities the mean free path is larger than the particle diameter, leading to a decreased drag
force due to the reduced number of collisions. This is taken into account by the Cunningham
slip-correction factor $C_\text{c}$~\cite{Hutchins:AST22:202}, which gives the drag force as
\begin{align}
  F_{\text{drag}} &= \frac{3\pi\mu d_{\text{p}}(\vec{U}-\vec{u})}{C_\text{c}} \label{eq:fdrag} \\
  \text{with } C_\text{c} &= 1+\Knud_\text{p}(c_1+c_2\cdot e^{c_3/\!\Knud_\text{p}})
  \label{eq:cc}
\end{align}
Here, $\vec{U}$ is the local velocity of the flow field, $\vec{u}$ the particle velocity and
$d_\text{p}$ the particle diameter. The empirical coefficients $c_1=1.2310$, $c_2=0.4695$ and
$c_3=-1.1783$ are taken from the literature~\cite{Hutchins:AST22:202}. This model describes the
interaction of a particle with a continuum flow field. In reality, however, the particles interact
\emph{via} single collisions with the carrier gas. This leads to diffusion and an
   additional random walk of the particles around their trajectory. This is numerically described
by a Brownian-motion force $F_{\text{b}}$, which is added to the drag force~\cite{Li:AST16:209}
\begin{equation}
   \label{eq:fb} F_\text{b}=m_{\text{p}}\vec{G}\sqrt{\frac{\pi S_0}{\Delta t}} \quad \text{with}
   \quad S_0=\frac{216\mu{k}T}{\pi^2d_\text{p}^5\rho_\text{p}^2C_\text{c}},
\end{equation}
where $\vec{G}$ is a vector of zero mean, unit variance, independent Gaussian random numbers,
$\Delta t$ the time step size of the solver, $k$ the Boltzmann constant, $T$ the temperature of the
carrier gas, $m_\text{p}$ the particle mass, $d_\text{p}$ the particle diameter and $\rho_\text{p}$
the particle density. Since the flow field was treated axisymmetrically, the Brownian force was
restricted to have axial and radial components. Particle trajectories are calculated
   until they reach the boundary of the flow field. This happens either when they are successfully
   transmitted to the end of the flow field downstream of the geometry or when they touch the wall
   of the geometry and are considered lost due to impaction.

\subsection{Experimental Setup}
\label{sec:exp_setup}
\begin{figure}
   \centering
   \includegraphics{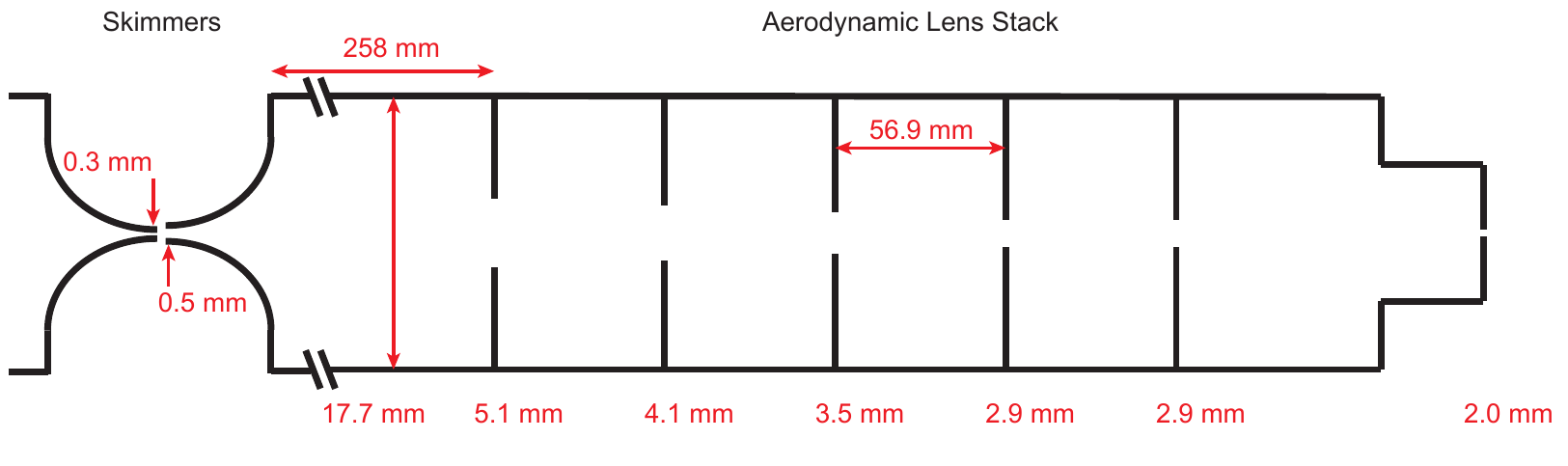}
   \caption{Sketch (not to scale) of the double skimmer setup and the ALS injector with its 6
      orifices. The dimensions given below the orifices refer to the inner diameter.}
   \label{fig:uppsalalens}
\end{figure}
To validate our simulations and the ability to predict ALS behavior, we benchmarked them against
experimental data. The experimental setup and data analysis has been described in detail
previously~\cite{Awel:OE24:6507}. Briefly, we used a gas dynamic virtual nozzle to aerosolize an
aqueous solution of 500~nm sized polystyrene spheres (Thermo Fisher Scientific) into a nebulization
chamber~\cite{DePonte:JPD41:195505,Beyerlein:RSI86:125104}. Particles then passed a set of two
skimmers, with inner diameters of 0.3~mm and 0.5~mm, respectively, placed 2~mm apart, as sketched in
\autoref{fig:uppsalalens}. Evacuating the volume between the skimmers allows control over the
pressure before the ALS while minimizing particle losses. After the ALS particles were illuminated
with a Nd:YLF laser (Spectra Physics Empower ICSHG-30, 527~nm, pulse duration 100~ns, pulse energy
20~mJ) and scattered light collected on a translatable high-frame-rate CMOS camera (Photron SA4)
using a $5\ordtimes$ infinity-corrected objective (Mitutoyo, numerical aperture 0.14). The
geometric dimensions of the ALS used are specified in \autoref{fig:uppsalalens}.

\subsection{Theoretical description of the experimental setup}
\label{sec:sim_setup}
For the theoretical model, we first calculated flow fields within which particles are then
propagated. When trying to simulate the entire apparatus, containing differential-pumping skimmers
and the ALS, we encountered convergence problems and no steady solution was found. Therefore, we
retreated to evaluate the flow fields for the ALS and the skimmer setup separately, but made sure
that they are consistent.

The effect of the skimmers was approximated by simulating the flow field through the upper skimmer,
simulating particle trajectories and retaining only particles with a radial position smaller than
250~\um at a position 2~mm downstream of the skimmer tip, representing those that would enter into
the second skimmer. The purpose of the upper skimmer is to accelerate particles, such that their
momentum is high enough to enter the lower skimmer without being significantly disturbed by the flow
field between the skimmers, where excess gasload is evacuated. Boundary conditions constrained the
inlet mass flow through the skimmer to 30~mg/min, comparable to experimental conditions. The outlet
was defined as a semi-circle at the tip of the skimmer with a 2~mm radius, corresponding to the
distance between the skimmers. Along this semicircle the pressure was constrained to experimentally
measured values. Particles were assumed to be spheres with 500~nm diameter and a density of
1050~kg/m$^3$ (polystyrene), with an initial uniform distribution at the entrance plane of the
skimmer. The longitudinal and radial velocities of the particles are set to the flow velocity at
their initial position. The recorded final phase-space distributions of transmitted particles are
used to define the initial particle phase-space distribution at the ALS.

To simulate the ALS we introduced boundary conditions for the pressure at the inlet and outlet. The
former was defined as the entrance plane at the beginning of the ALS tube, and pressures set to
experimental values. The outlet was defined as a semi-circle at the tip of the ALS into vacuum with
radius 1~mm, corresponding to the radius of the final aperture of the ALS. The pressure along this
semicircle was assumed to be $10^{-2}$~mbar. Reducing this pressures further does not change the
dynamics in the flow, since they depend on pressure difference, which is already dominated by the
two-orders-of-magnitude higher pressure inside the ALS. With these boundary conditions we calculated
a steady flow field for every inlet pressure. The initial phase-space distributions of particles
were taken from the skimmer simulations, but with the initial longitudinal position of all particles
set to the entrance plane at the beginning of the ALS tube.

We simulated $10^5$ particles per upstream pressure. Final particle trajectories contain the
axisymmetric two dimensional position of particles throughout and after the ALS. In the experiment,
we probed the particle beam orthogonal to the propagation direction by projecting it onto the
imaging plane of a camera~\cite{Awel:OE24:6507}. Hence, the simulated radial particle beam
distribution was projected \emph{in silico} onto a two dimensional imaging plane for comparison.

\section{Results \& Discussion}
\label{sec:results}
\subsection{Validation against literature simulations}
\label{sec:comp_literature}
\begin{figure}[b]
   \centering
   \includegraphics[width=0.75\linewidth]{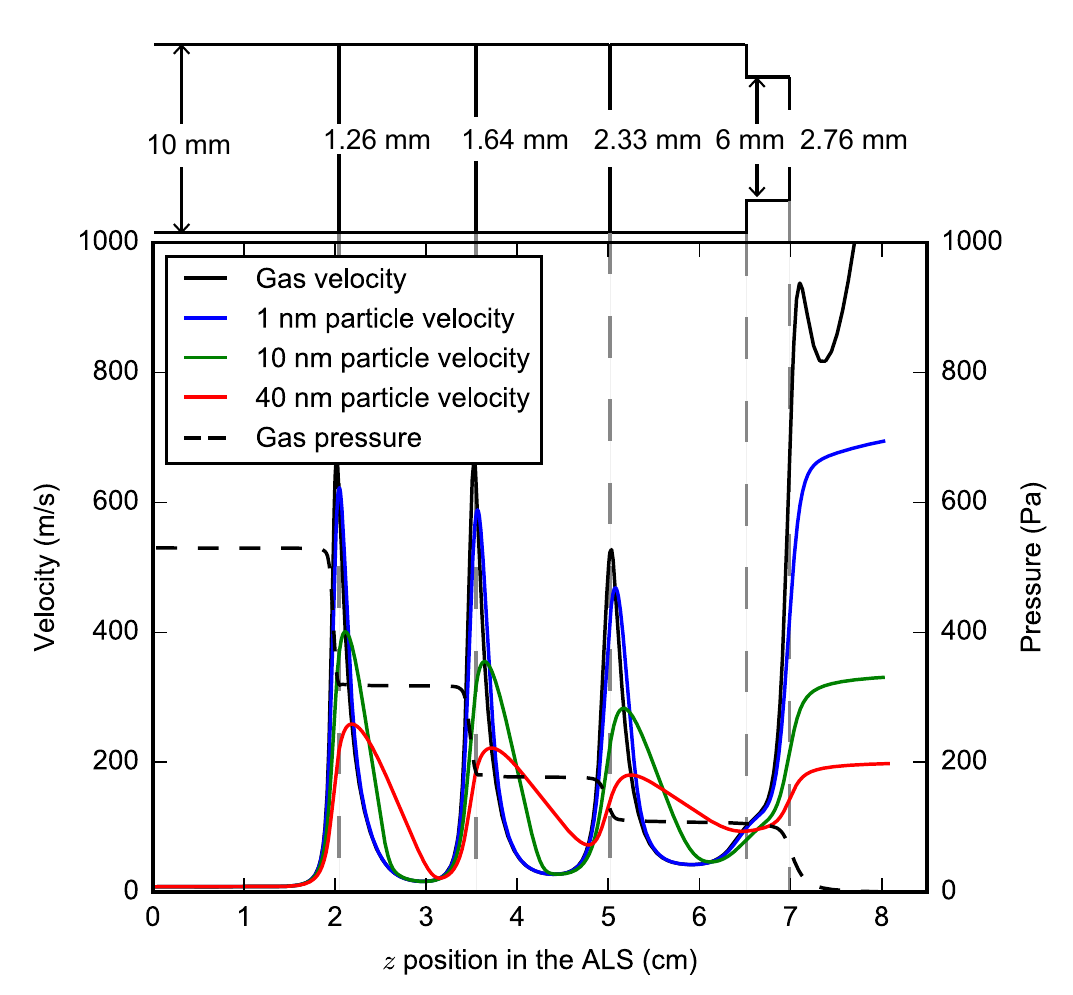}
   \caption{Simulation of gas pressure, gas velocity and particle velocity of $1$, $10$ and 40~nm
      particles along the centerline of the ALS investigated in~\cite{Wang:AST39:624}, and shown at
      the top of the figure. Our simulation accurately reproduced the results from the original
      publication~\cite[Figure~2]{Wang:AST39:624}.}
   \label{fig:wang2005}
\end{figure}
We first validated our simulation environment against the simulations by Wang
\etal~\cite{Wang:AST40:320}. We replicated the geometry and conditions of the original publication
and simulated particle trajectories. \autoref{fig:wang2005} shows the local gas velocity and
pressure as well as the velocity of $1$, $10$ and $40$~nm particles along the centerline of the ALS.
Our results show excellent agreement with the previously published simulations, \cf Figure~2b of
reference~\citealp{Wang:AST39:624}, and only deviate slightly in the region outside the actual ALS
($z>7$~cm). These deviations can be explained by the different outlet boundary conditions.

\subsection{Particle Beam Characterization}
\subsubsection{Experimental Results}
\begin{figure}[t]
   \centering
   \includegraphics{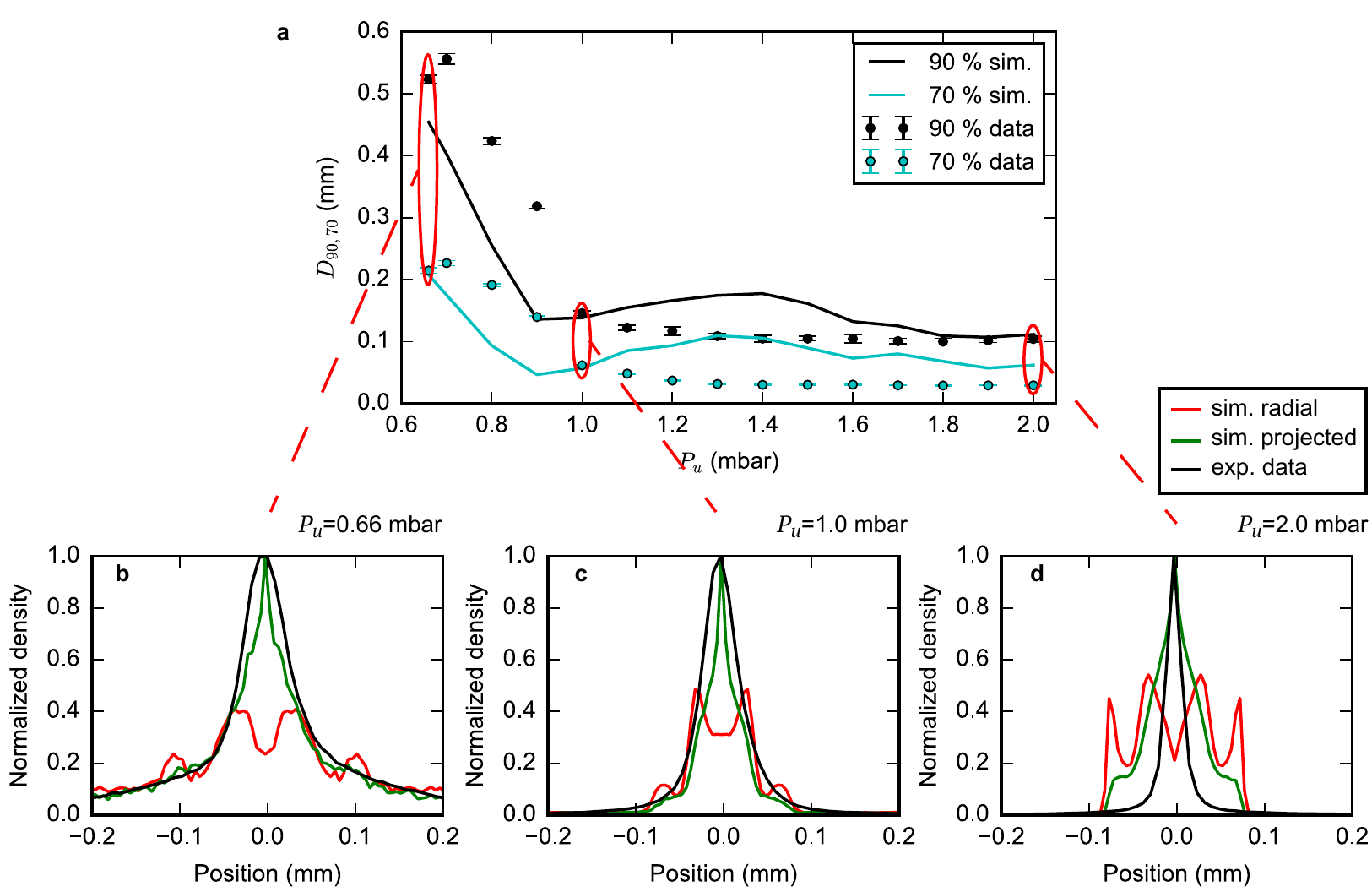}
   \caption{Comparison of experimental, \ie, projected, and simulated, \ie, projected \emph{in
         silico}, beam width for 500~nm particles as a function of inlet pressure $P_u$. The
      errorbars represent the statistical standard error (\textbf{a}). Comparison of
         experimental, \ie, projected (black line), simulated radial (red line), and simulated
         projected (green line),} particle profiles for three distinct upstream pressures of the
      ALS, 0.66 mbar (\textbf{b}), 1.0~mbar (\textbf{c}) and 2.0~mbar (\textbf{d}).
   \label{fig:sim_res}
\end{figure}

All measurements were conducted with the setup described in \autoref{sec:exp_setup}, and the only
parameter varied was the pressure upstream of the ALS. The particle stream was imaged 8~mm
downstream of the injector tip and data collected for $\ordsim10$~min at each pressure,
corresponding to $\ordsim10^5$ imaged particles. For comparison with theoretical results, we
determined the particle beam width containing 90~\% (70~\%) of all particles, denoted $D_{90}$
($D_{70}$). Measured beam widths are shown in \subautoref{fig:sim_res}{a} for upstream pressures in
the range 0.66 to 2.0~mbar.

The full distributions for three characteristic pressures are shown in
\subautoref{fig:sim_res}{b--d} (black lines). It is evident from the experimental data that the
particle beam width decreases with increasing upstream pressure until a critical value, here
$\ordsim1.2$~mbar, after which no dependence on pressure is observed anymore and the produced beam
width remains practically constant.

\subsubsection{Theoretical Results}
\label{sec:theo_res}
\begin{figure}
   \centering
   \includegraphics{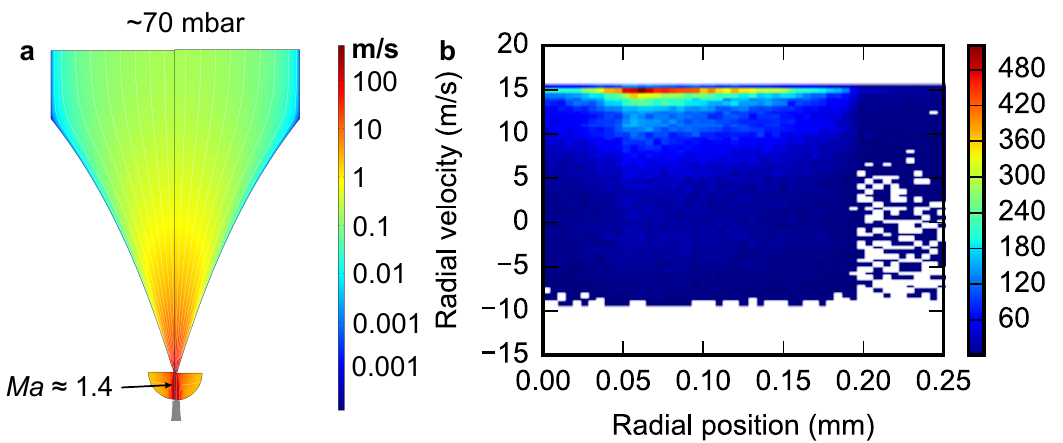}
   \caption{Simulated flow field with streamlines inside the upstream skimmer. The logarithmic color
      scale corresponds to the flow speed. The grey area indicates the position of the lower skimmer.
      (\textbf{a}). Histogram of the radial position and velocity of particles 2~mm downstream from
      the skimmer tip (\textbf{b}).}
   \label{fig:skimmer}
\end{figure}

In order to simulate the resulting particle beam downstream of the ALS, it is necessary to compute
the initial phase-space distribution of particles entering the ALS by simulating the first skimmer
as detailed above. An example of a velocity field inside and such a radial phase-space distribution
of particles after the first skimmer is shown in \subautoref{fig:skimmer}{a} and b, respectively.
The radial position is cut at 250~\um, as only these particles enter into the second skimmer. A
large fraction of particles is contained within a small region of phase space at radial velocities
between 10 and 15~m/s.

\begin{figure}
   \centering
   \includegraphics{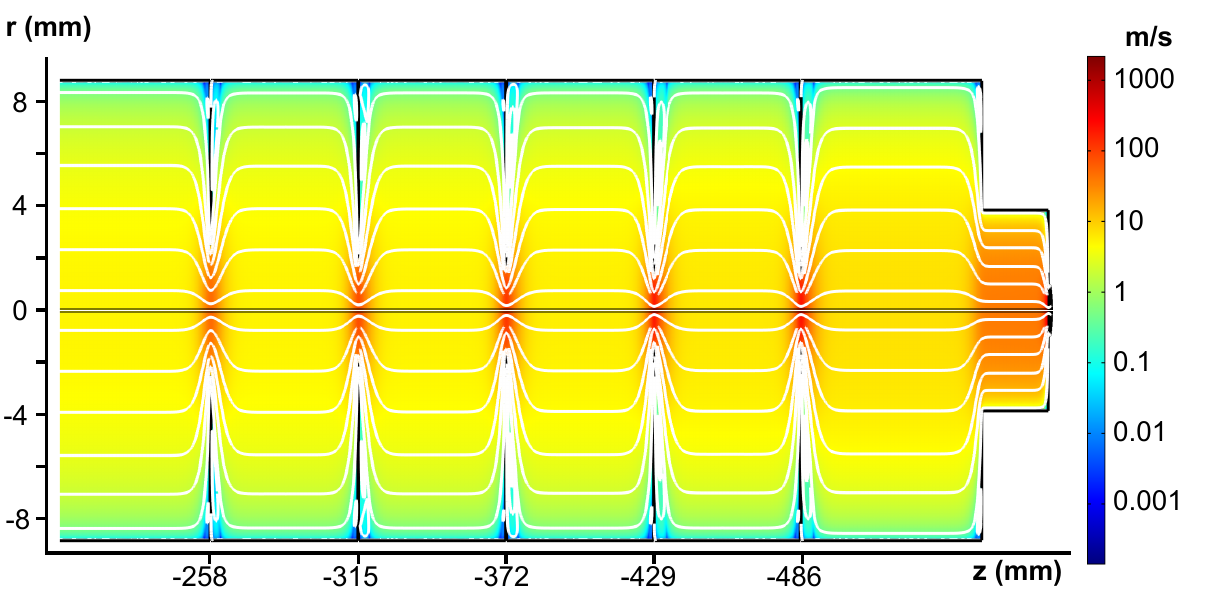}
   \caption{Simulated flow field with streamlines inside the ALS injector. The $r$-axis has been
      scaled by a factor of 10. The logarithmic color scale corresponds to the flow speed.}
   \label{fig:flowfield}
\end{figure}
\begin{figure}
   \centering
   \includegraphics[width=0.95\linewidth]{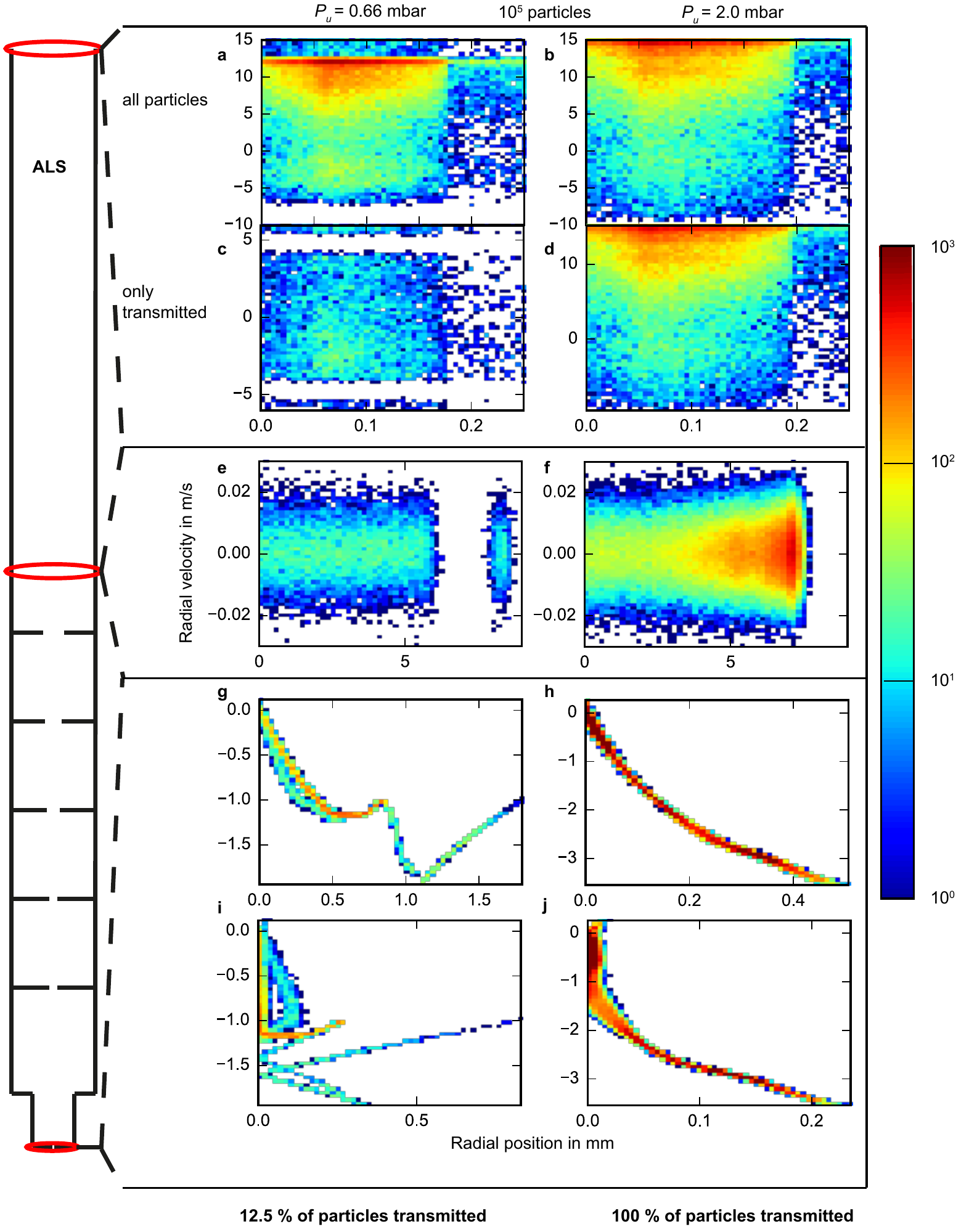}
   \caption{Histograms of the radial phase-space distribution for 500~nm particles at various
      positions in the ALS for $P_u$=0.66~mbar (\textbf{a, c, e, g, i}) and $P_u$=2.0~mbar
      (\textbf{b, d, f, h, j}). \textbf{a, b} show the distributions for all particles at the inlet
      of the ALS. All other distributions show only particles successfully transmitted through the
      ALS and are taken at a position at the inlet (\textbf{c, d}), before the first lens
      (\textbf{e, f}), after the ALS (\textbf{g, h}) and at the distance of highest density, 10~mm
      (\textbf{i}) and 5~mm (\textbf{j}) downstream of the outlet, as indicated by the red lines.}
   \label{fig:2d}
\end{figure}

Simulated radial phase-space density distributions of particles through the flow-field shown in
\autoref{fig:flowfield} at various positions within the ALS are displayed in \autoref{fig:2d} for
two different upstream pressures.
The left hand side of \autoref{fig:2d} corresponds to 0.66~mbar and the right hand side to 2.0~mbar.
The top images show the initial distribution (beginning of the injector), while all others contain
only successfully transmitted particles, which are shown at three positions; at the beginning of the
ALS (c and d), just before the first ADL (e and f), 1~mm after exiting the ALS (g and h) and at the
particle focus 10~mm and 5~mm after exiting the ALS, respectively (i and j). Whereas the initial
phase-space distributions of particles are nearly identical for the two pressures, the number of
particles transmitted and their phase-space distributions throughout and after the ALS are markedly
different. While for high $P_\text{u}$ all particles are transmitted, this is not the case at low
$P_\text{u}$. Here, the initial phase-space distribution of successful transmitted particles
exhibits a cutoff at around 6~m/s absolute radial velocity (\subautoref{fig:2d}{c}). We rationalize
this with the pressure scaling of the drag force. Particles at lower pressures, and hence lower drag
force, are not slowed down sufficiently in the radial direction, such that they collide with the
wall of the ALS tube and are lost. This is not the case at $P_\text{u}=2$~mbar, where all particles
are slowed down before they reach the wall and are transmitted. The radial velocity of the particles
before the first lens is now essentially zero and most particles lie within $\pm0.02$~m/s. The
phase-space densities in \subautoref{fig:2d}{e} and f show that particle radial positions are spread
over the entire ALS tube, and that at higher pressures (f) the majority of particles are at large
radii. These particles at large radial positions correspond to those with an initially large radial
velocity. In the low pressure case (e) high-radial-velocity particles collided with the wall and
were lost. This correlation between initial radial velocity and radial position is, furthermore,
evident for the low pressure case (\subautoref{fig:2d}{c} and e), where an initially empty area of
velocity space (between 4 and 5~m/s) appears as an empty area in position space (around 7~mm) before
the first ADL. Thus the acceptance of the ALS depends on the upstream pressure, \ie, the flow field,
and in the low pressure case the transmission and behavior of the ALS depend critically on the
radial particle position before the first ADL, hence the initial radial particle velocity. This
position-dependent behavior of an ADL will be further investigated in
\autoref{sec:radial_distribution_analysis}.

In both pressure regimes the distribution of final radial positions (\subautoref{fig:2d}{g} and h)
is concentrated toward the centerline in comparison to the distribution before the first lens
(\subautoref{fig:2d}{e} and f) and the final radial velocity distribution is narrower, hence, the
particle beams are more collimated than the inlet distributions (\subautoref{fig:2d}{a} and b).
However, in the high pressure case particles are confined to significantly smaller radii
corresponding to $D_{70}=226$~\um at this position in comparison to $D_{70}=815$~\um for the low
pressure case. The particles radial velocity is predominantly negative after the lens, corresponding
to a motion toward the centerline, \ie, the particle beam converges. Particles at higher radial
positions have a greater negative radial velocity and, therefore, a higher density of particles will
be achieved downstream of the injector. In the 0.66~mbar case the highest density is achieved 10~mm
after the injector outlet, while for 2~mbar it is 5~mm downstream of the outlet. The corresponding
phase-space distributions are shown in \subautoref{fig:2d}{i} and j. Note that from 1~mm downstream
of the injector (the end of the calculated steady-state flow field) onwards the particles are
propagated straight without any forces acting on them.

\begin{figure}
   \includegraphics{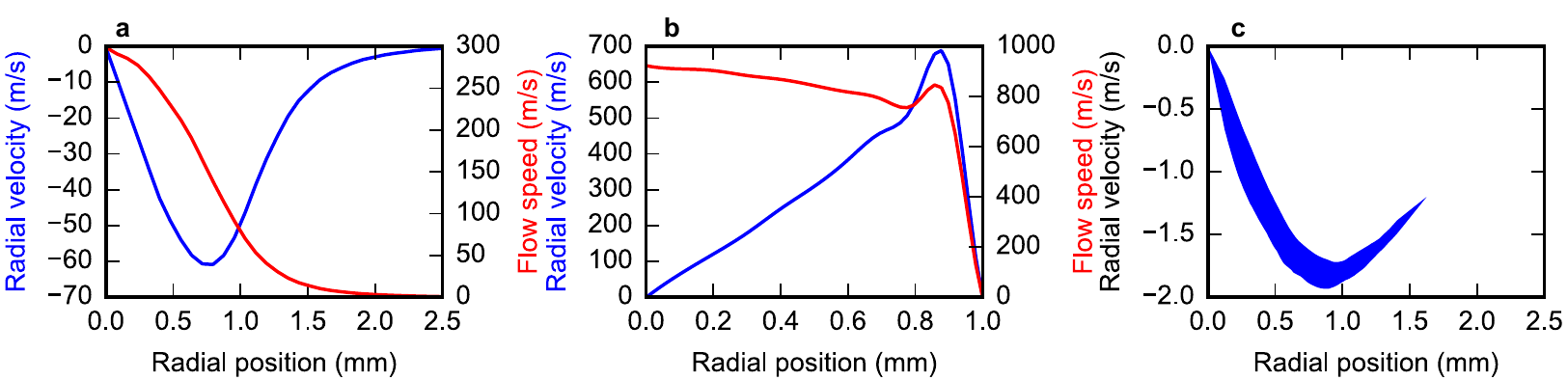}
   \centering
   \caption{Radial velocity and speed of the helium flow as a function of radial position for
      0.66~mbar upstream pressure at a position 0.2~mm before (\textbf{a}) and 0.5~mm after
      (\textbf{b}) the last orifice in the ALS injector. Phase-space distribution of 500~nm
      particles 0.2~mm before the last orifice in the ALS injector at a 0.66~mbar upstream pressure
      (\textbf{c}).}
   \label{fig:explanation}
\end{figure}

The final phase-space distribution is predominantly defined by the last aperture, as the radial
velocity upstream of each ADL is centered around 0, see \subautoref{fig:2d}{e} and f. In order to
qualitatively rationalize the observed distributions we consider the radial-position dependence of
the radial velocity and total speed of the gas flow before (\subautoref{fig:explanation}{a}) and
after (\subautoref{fig:explanation}{b}) the last lens in the ADL. The radial velocity of the flow
changes throughout the orifice, from a contraction towards the centerline, \ie, negative radial
velocities in \subautoref{fig:explanation}{a}, to an expansion afterwards, positive radial velocity
in \subautoref{fig:explanation}{b}. This dramatic change is caused by the significantly different
pressure regimes, inside the ADL \emph{versus} outside the ADL. Since the drag force is proportional
to the difference in particle velocity and flow field velocity, see \eqref{eq:fdrag}, the force
acting on a particle in the radial direction is proportional to the radial flow velocity. In
\subautoref{fig:explanation}{c} we show the particle phase-space distribution before the last
orifice, \ie, at the same position as the gas flow distribution shown in
\subautoref{fig:explanation}{a}. A clear correlation is observed between the radial gas-flow
velocity, blue line in \subautoref{fig:explanation}{a}, and the particle phase-space distribution.
As the radial velocity of the gas is changing rapidly on passing the orifice, one might expect a
similar effect on the radial velocity distribution of particles. However, since particles carry a
significant amount of inertia, they cannot follow this rapid change in gas-flow and the particle
phase-space distribution even after the last aperture (as shown in \subautoref{fig:2d}{g}) is still
dominated by the distribution \emph{before} the orifice. One noticeable difference, however, is an
increase of radial velocity around the position 0.8~mm. We attribute this to the rapid expansion of
gas after the last aperture, which peaks at this radial position (\subautoref{fig:explanation}{b})
and, hence, accelerates particles most at this distinct radius, leading to the observed local
maximum in the particle radial velocity around 0.8~mm.

Where within the general shape of the phase-space distribution particles are located (\ie, the intensity
information missing in \subautoref{fig:explanation}{c}) is dependent on the radial position of
particles upstream of an ADL. Since, in the high pressure case, no particles are at
radial positions above $\ordsim0.5$~mm, only the initial falling edge at small radial positions is
represented in the phase-space distribution in \subautoref{fig:2d}{h}.

\subsubsection{Comparison of Simulation and Experiment}
\label{sec:sim_v_exp}
We compared the simulated results with experimental data by reproducing the pressure dependence of
the particle-beam width evaluated 8~mm downstream of the ALS tip, as shown in
\subautoref{fig:sim_res}{a}. The simulations clearly reproduced the experimental observation, with a
sharp drop in beam diameter as the pressure is increased, until a plateau is reached at
$\ordsim0.9$~mbar. Full radial distributions of particles are shown in
\subautoref{fig:sim_res}{b--d} for three inlet pressures and exhibit an overall good agreement with
the experimental data. Some deviations are, however, observed. A slight pressure offset for the
location of the sharp drop is most likely due to the $\pm15~\%$ uncertainty of the pressure gauges
used (Pfeiffer Vacuum, TPR 280). Moreover, the simulations overestimate the particle beam size in
the plateau region, which could be due to the limited illumination area of the laser used for
particle detection~\cite{Awel:OE24:6507}. If particles far from the center were not correctly
identified, this would lead to lower than expected experimental values for $D_{70}$ and $D_{90}$. On
the other hand, for very high particle densities, there is a probability that the image analysis
software cannot distinguish individual particles anymore. This would lead to a decreased particle
density detected in the central region.

We also note that our simulation might oversimplify the occurring physical processes, \eg, particles
are assumed to have no collisions 1~mm downstream of the last aperture, whereas there are some
experimental indications that particles still accelerate in this
region~\cite{Bielecki:privcomm:2017}. In addition, the used Cunningham correction factors, see
\autoref{sec:particle-traces}, were derived for air instead of helium. While this might render it
difficult to computationally reach high accuracy, the overall good agreement justifies the use of
these models, and our simulation infrastructure in general, to understand and predict ALS behavior.

\subsection{Radial distribution analysis}
\label{sec:radial_distribution_analysis}
In SPI experiments the interaction volume is a cylinder, representing the x-ray beam volume, through
the three dimensional particle beam. Therefore, the vast majority of particles with a radial
position greater than the x-ray spot will not interact with the photons. The radial particle
distributions in \subautoref{fig:sim_res}{b--d} (red lines) show that, additionally to the main
peak, smaller outer secondary maxima are present, which reduce the number of particles contained
within the interaction volume. The formation of these ``wings'' has also been observed in other
studies of particle distributions from ALS~\cite{Giuseppe:AST33:105, Headrick:JAS58:158}, but no
explanation as to the source of this effect was given.
\begin{figure}
   \centering
   \includegraphics{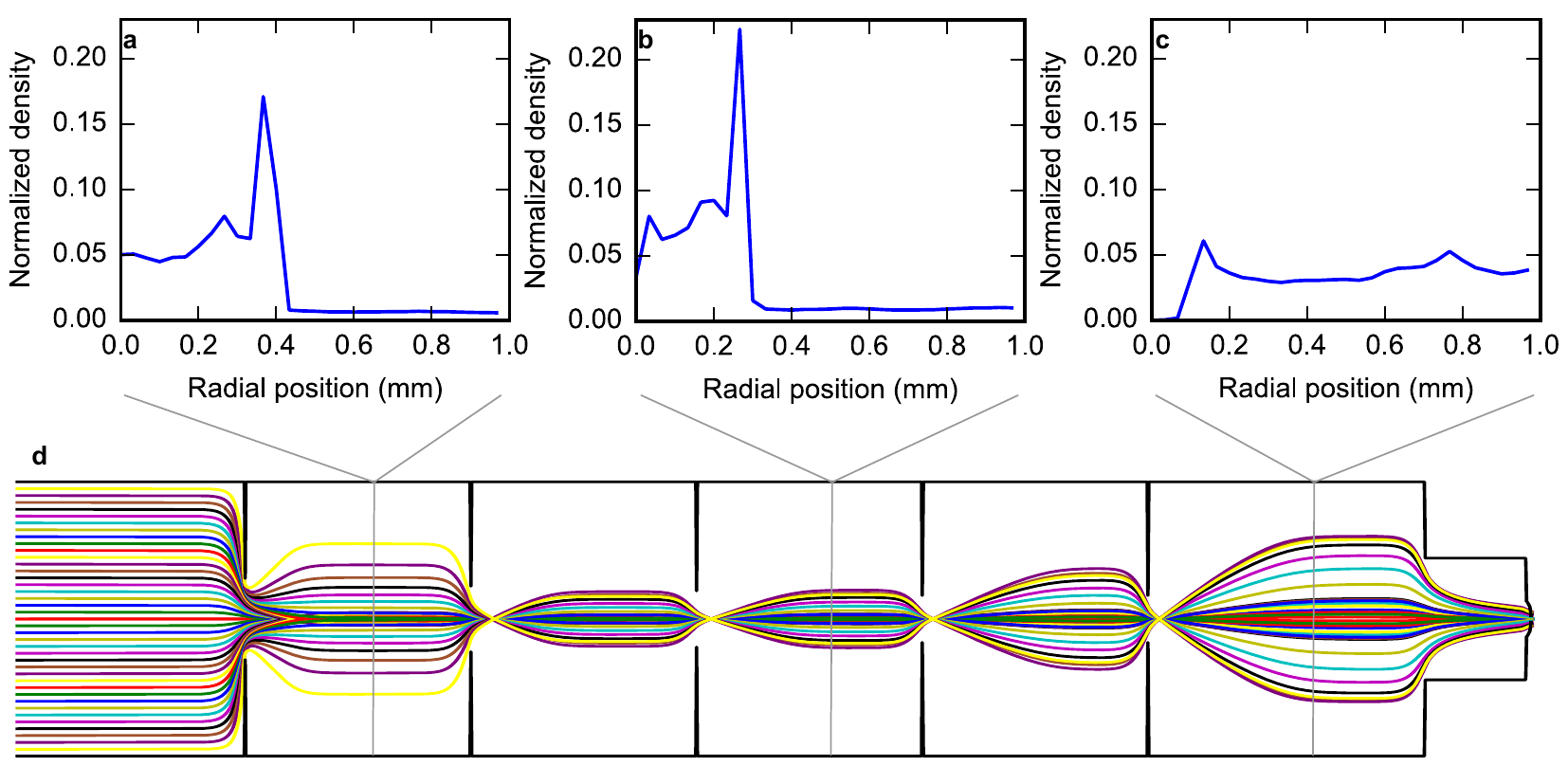}
   \caption{Radial distributions of 500~nm particles at different stages inside the ALS injector for
      a uniform initial radial distribution (\textbf{a-c}). The formation of a peak at large radii
      (\ie, a ``wing'' in the particle distribution) is already evident after the first lens
      (\textbf{a}). Example trajectories of 500~nm particles through the ALS injector, without
      diffusion effects (\textbf{d}).}
   \label{fig:particle_distributions_inside_uppsala}
\end{figure}

\begin{figure}[t]
   \includegraphics{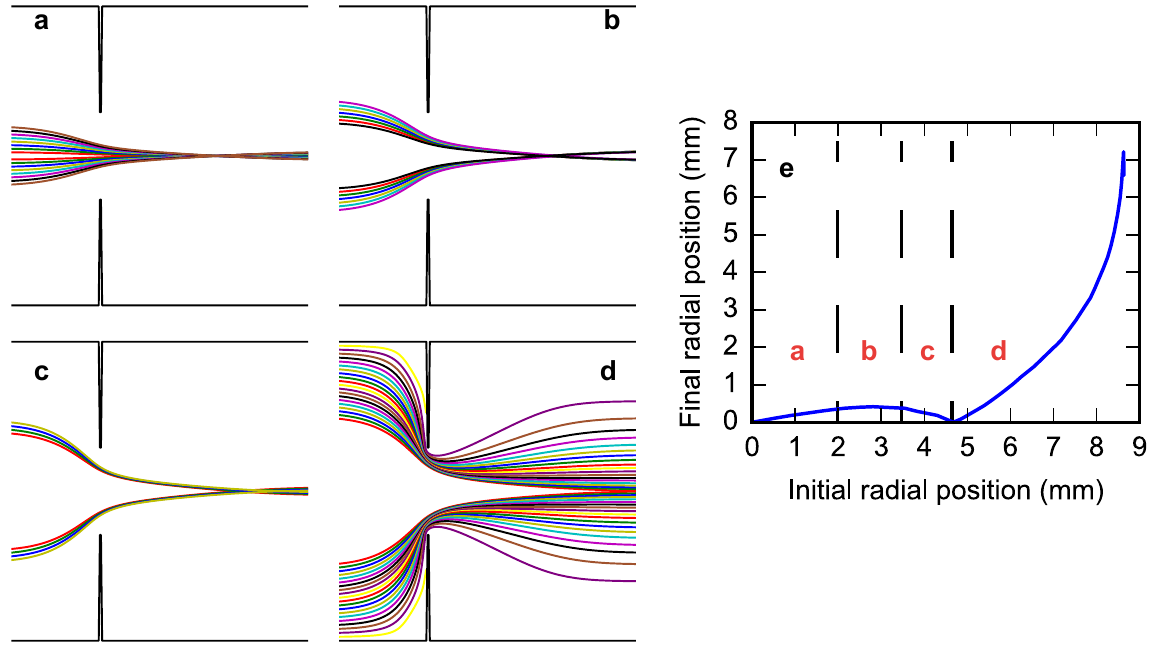}
   \centering
   \caption{Trajectories of 500~nm particles through the first lens of the ALS injector, for
      different radial starting positions of $0-2$~mm (\textbf{a}), $2-3.5$~mm (\textbf{b}),
      $3.5-5.0$~mm (\textbf{c}) and $5.0-8.8$~mm (\textbf{d}). Correlation between initial and final
      radial position for particles traveling through the first lens (\textbf{e}). We identify four
      distinct regions, as exemplified by the trajectories in \textbf{a--d}.}
   \label{fig:initial_final}
\end{figure}

To investigate this, we considered the radial distributions of particles within the ALS and show
these after the first, third, and fifth lens in
\subautoref{fig:particle_distributions_inside_uppsala}{a--c}, respectively. These were simulated for
2~mbar inlet pressure and 0.31~mbar downstream pressure, with particles evenly distributed at the
inlet and neglecting Brownian motion. Example trajectories through the entire ALS are shown in
\subautoref{fig:particle_distributions_inside_uppsala}{d}. It is immediately evident that an outer
maximum in the radial distribution is already present after the first lens, and that not all lenses
are contracting the particle beam, with some even broadening the distribution. These effects are due
to the design of this ALS to accept a large range of particle sizes. In order to visualize the
origin of the outer radial maxima we considered particle trajectories through the first lens for
different radial starting positions, \autoref{fig:initial_final}.

A particle exactly on the centerline of the ADL simply stays there, as it feels no radial force, see
\subautoref{fig:explanation}{a}. The further off-center the particle is located, the larger the
curvature of the flow toward the centerline, leading to a larger radial force. This leads to
particles with initial radial positions between 0--2~mm (\subautoref{fig:initial_final}{a}) getting
pushed towards the center. Due to their inertia particles cross the centerline, but remain closer to
it than initially, \ie, the beam is contracted and larger initial radial positions lead to larger
final radial positions.

Further away from the centerline the curvature of the flow still increases, while flow speed
decreases with proximity to the outer wall (see \subautoref{fig:explanation}{a}). These
counter-acting mechanisms negate each other for particles with initial radial positions between 2
and 3.5~mm (\subautoref{fig:initial_final}{b}) and in this region all particles arrive at
approximately the same final radial position, regardless of their initial radial position.

Increasing the initial radial position even further ($3.5$ to 5.0~mm,
\subautoref{fig:initial_final}{c}) leads to the decreasing flow speed dominating and final radial
positions get closer to the centerline with increasing initial radial position, \ie, the opposite
effect to that observed in \subautoref{fig:initial_final}{a}.

Eventually, at around 5.0~mm (\subautoref{fig:initial_final}{d}), trajectories stop crossing the
centerline and the final radial position increases with increasing initial radial position again.
This overall behavior is also summarized in \subautoref{fig:initial_final}{e}, showing the radial
position after the first lens as a function of the initial radial position. The secondary maxima
observed in the radial distribution in \autoref{fig:sim_res} thus arise at the turning point
in b, where several initial radial positions result in the same final position, hence leading to
an increased particle density at distinct radii.

This undesirable behavior can be mitigated by designing an ALS such that it only
operates in either one of the regimes corresponding to \subautoref{fig:initial_final}{a} or d.
Total avoidance of secondary maxima can only be accomplished by operating
exclusively in regime a, which would -- conceptionally -- be the best
solution to this problem. However, it is experimentally impractical, because it requires an ALS
tube much larger than the radial size of the incoming particle beam. Designing a lens such that
crossing of the centerline is minimized would ensure that more particles, including
those at small initial radii, will be in the regime d, producing a more collimated particle stream
at the output, with fewer particles in secondary maxima, but also with a reduced
amount of focusing. This can be achieved by increasing the orifice diameters of the ADL for a given
mass flow. This way the absolute value of the derivative of the radial velocity of the gas flow
before the orifice with respect to the radial position decreases, while the flow speed drop caused
by the walls remains the same. Hence, the radial velocity minimum has an increased value and its
position changes to smaller radii, \ie, the minimum in \subautoref{fig:explanation}{a} moves to the
upper left. The regimes a--c are moved towards smaller radii until they get negligible.
\begin{figure}[t]
   \centering
   \includegraphics{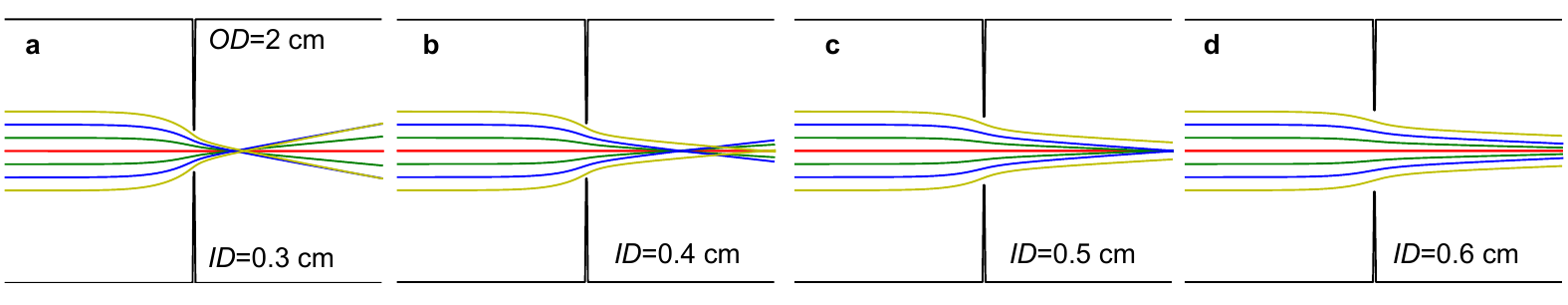}
   \caption{Simulated trajectories of 500~nm particles through an ADL with an outer diameter of 2~cm
      and an inner diameter of 3~mm (\textbf{a}), 4~mm (\textbf{b}), 5~mm (\textbf{c}), 6~mm
      (\textbf{d}). The pressure downstream of the ADL is 0.5~mbar and the mass flow of carrier gas
      is $1.2\cdot10^{-2}$~mg/min.}
   \label{fig:optimizing_lens}
\end{figure}
This is highlighted in \autoref{fig:optimizing_lens}, showing particle trajectories through lenses
with various inner diameters for an identical mass flow. Increasing the orifice diameter shifts the
crossing point of trajectories further away from the lens and a more collimated particle stream is
produced (\subautoref{fig:optimizing_lens}{d}). Thus operation in regime d is readily
   achievable, but the corresponding effects of stronger collimation and weaker focusing requires
   more ADLs for reaching high densities. Balancing these effects for a limited amount of ADLs in
   the ALS, secondary maxima will not completely be avoided when maximizing the central particle
   density.

\begin{figure}
   \centering
   \includegraphics{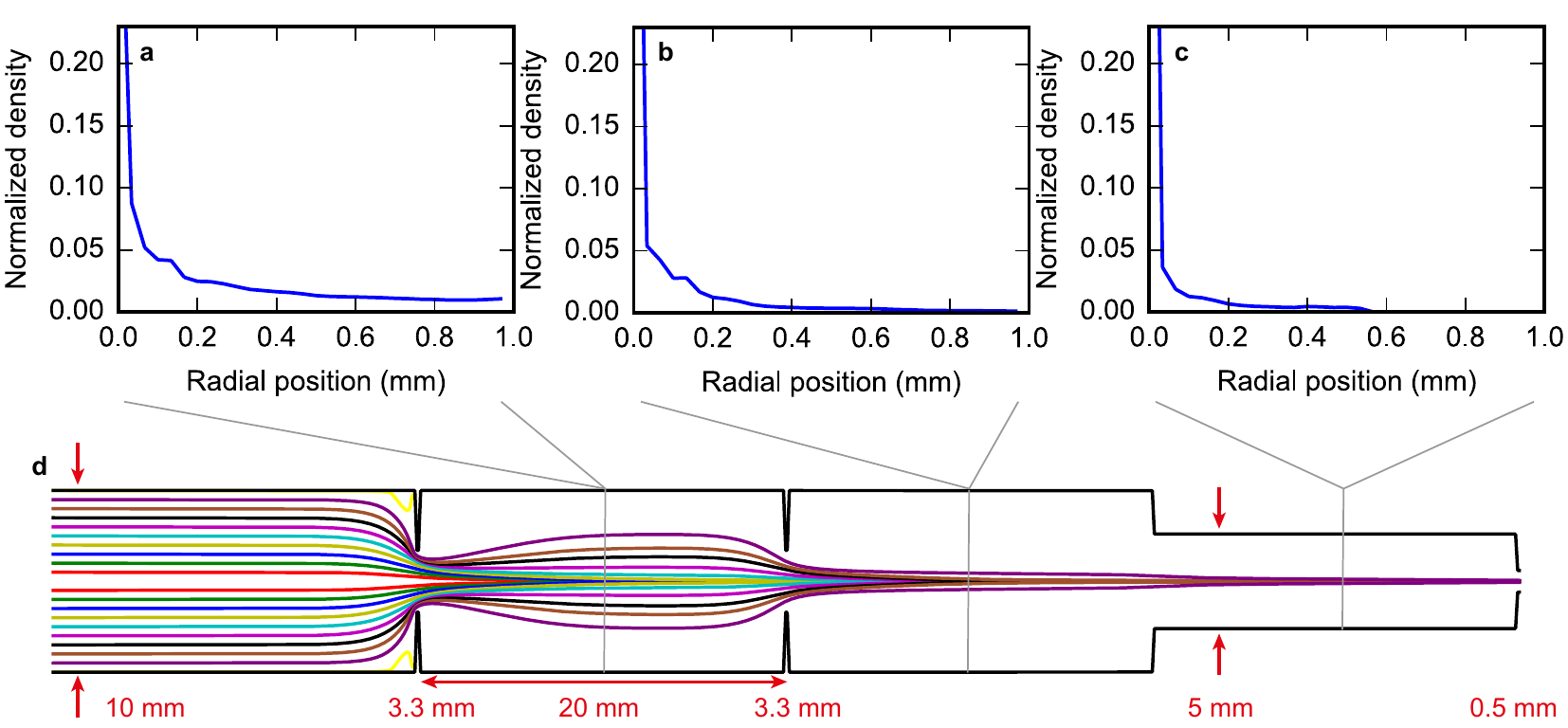}
   \caption{Radial distributions of 500~nm particles at different stages inside an optimized ALS
      injector (\textbf{a-c}). The formation of peaks at large radii (\ie, a ``wing'' in the
      particle distribution) is significantly reduced in comparison to the conventional ALS in
      \autoref{fig:particle_distributions_inside_uppsala}. Example trajectories of 500~nm particles
      through the optimized ALS injector, without diffusion effects (\textbf{d}).}
   \label{fig:particle_distributions_inside_new_injector}
\end{figure}
\autoref{fig:particle_distributions_inside_new_injector} shows a three-lens-system for focusing
500~nm particles that way. \subautoref{fig:particle_distributions_inside_new_injector}{a--c} shows
radial distributions of 500~nm particles at various positions within the new ALS, demonstrating that
particles are smoothly collimated toward the centerline with significantly weaker
   secondary maxima than for the conventional ALS in
   \autoref{fig:particle_distributions_inside_uppsala}.
\begin{figure}[t]
   \centering%
   \includegraphics{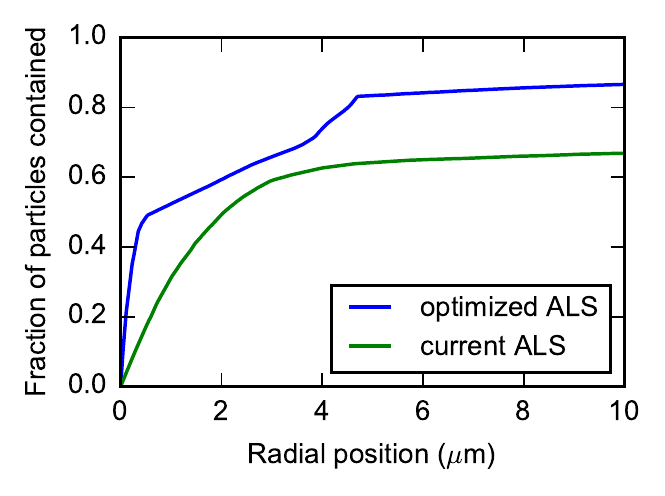}
   \caption{Fraction of particles within a given radius at the position of the
         respective smallest particle beam waist for the ''Uppsala'' injector (4.6~mm downstream the
         exit) and the new ALS design (0.5~mm downstream the exit).}
   \label{fig:old_vs_new}
\end{figure}
To compare the particle beams of the optimized and the ''Uppsala'' ALS, we evaluated the fraction of
particles arriving within a given radius at the respective particle beam focii, 4.6~mm
and 0.5~mm downstream for the "Uppsala" and optimized injector, respectively; see
\autoref{fig:old_vs_new}. The optimized ALS exhibits a much steeper increase of the integrated
particle fraction at small radii, \ie, focusing a significantly larger fraction of particles into a
given radius. As discussed above, secondary maxima cannot be avoided completely in the
improved ALS, leading to a kink in the fraction of particles contained around a radius of
$\ordsim0.5~\um$. In SPI experiments, it is especially pertinent to compare the fraction of
particles that would cross the interaction volume with the x-rays. For an x-ray focal
spot size with a radius of 0.5~\um, in the current lens design only 17.1~\% of the particles are
contained within that radius, while in the optimized design this increases to 48.5~\% ---
corresponding to a nearly threefold improvement. At the same time, the final particle density in the
interaction volume also depends on the velocity of the particles leaving the injector. With lower
velocities the particles are ``packed'' closer in the z dimension and the density is correspondingly
higher. For the optimized injector the mean final particle velocity is 43.8~m/s, whereas for the
"Uppsala" injector it is 57.9~m/s. This leads to a velocity weighted density within the 0.5~\um
radius spot that is higher by a factor of $\ordsim4$ for the optimized injector; for an x-ray focal
spot size of 50~nm radius it would even be higher by a factor of $\ordsim9$. While the simulated
increase in particle density might not be quantitatively accurate, since this simulation was not
taking into account particle diffusion or the skimmer setup, which could significantly influence the
final particle beam, see \autoref{sec:simulation}, it is, nonetheless, clear that an optimized ALS
with a compact three-lens design can achieve significantly better particle beam concentration than
current injectors.

\section{Conclusion}
\label{sec:conclusion}
The results of a new computer-simulation environment for ALS injectors have been presented. Previous
theoretical treatments were quantitatively well reproduced. Focusing on the development of ALS for
SPI experiments, the priority is to maximize the particle density along the centerline of the
produced particle beam. Comparison of simulated particle profiles with experimental measurements
show a good agreement, further validating our computational approach and the ability to describe the
experimentally observed behavior. By computing particle trajectories through the ALS, our simulation
framework can provide a detailed insight into the particle dynamics inside the ALS, such as the
radial position dependent concentration mechanism of an ADL, and hence, the resulting particle
profiles. This way we were not only able to monitor, \eg, the overall pressure dependence of the
resulting particle beam, but to understand the mechanisms inside an ALS that are responsible for
specific artifacts in the radial particle distribution. We could pin down the source of the majority
of particle losses in current ALS to be caused by the double skimmer setup \emph{before} the first
lens.

Furthermore, we analyzed the formation of secondary maxima in the radial particle distribution in
current ALS and found these to be caused by particles crossing the axial centerline inside the ALS.
We demonstrated that it is feasible to design a simple ALS that avoids this problem altogether for
particles within a narrow size range. It produces a tightly focused stream of particles
exhibiting less secondary maxima and a significantly, nearly fourfold, increased
particle density at the center of the distribution. In an ALS designed for a wide range of particle
sizes the defocusing process and the trapping of particles in secondary maxima cannot be avoided.
Therefore, a simple ALS injector system, designed only for a specific particle size, is
better suited to fulfill the stringent requirements for atomic-resolution
single-particle diffractive imaging and other applications that require highest particle densities.
The quick exchange of lenses to adjust for distinct samples would be advantageous for
high-throughput experiments. Such an ALS setup is currently under development in our laboratory,
along with further quantitative measurements of particle and absolute gas densities emerging from
the injector~\cite{Awel:OE24:6507, Horke:JAP121:123106}, to benchmark and improve simulations by
comparison to experiment.

Furthermore, we point out that such an optimized ALS provides a spatial separation of different
species that might be present in the original aerosol, similar to more specific separation
techniques for small molecules~\cite{Filsinger:PRL100:133003, Chang:IRPC34:557}, and thus provides a
more homogeneous sample for SPI experiments~\cite{Barty:ARPC64:415}.

\section*{Acknowledgments}\noindent%
In addition to DESY, this work has been supported by the European Research Council under the
European Union's Seventh Framework Programme (FP7/2007-2013) through the Consolidator Grant COMOTION
(ERC-Küpper-614507), by the excellence cluster ``The Hamburg Center for Ultrafast Imaging --
Structure, Dynamics and Control of Matter at the Atomic Scale'' of the Deutsche
Forschungsgemeinschaft (CUI, DFG-EXC1074), and by the Helmholtz Gemeinschaft through the ``Impuls-
und Vernetzungsfond''.

\vspace{1em}

\bibliography{string,cmi}

\begin{thebibliography}{10}
\expandafter\ifx\csname url\endcsname\relax
  \def\url#1{\texttt{#1}}\fi
\expandafter\ifx\csname urlprefix\endcsname\relax\def\urlprefix{URL }\fi
\expandafter\ifx\csname href\endcsname\relax
  \def\href#1#2{#2} \def\path#1{#1}\fi

\bibitem{Bogan:NanoLett8:310}
M.~J. Bogan, W.~H. Benner, S.~Boutet, U.~Rohner, M.~Frank, A.~Barty, M.~M.
  Seibert, F.~Maia, S.~Marchesini, S.~Bajt, B.~Woods, V.~Riot, S.~P. Hau-Riege,
  M.~Svenda, E.~Marklund, E.~Spiller, J.~Hajdu, H.~N. Chapman,
  \href{http://pubs.acs.org/cgi-bin/abstract.cgi/nalefd/2008/8/i01/abs/nl072728k.html}{Single
  particle x-ray diffractive imaging}, Nano Letters 8~(1) (2008) 310--316.
\newblock \href {http://dx.doi.org/10.1021/nl072728k}
  {\path{doi:10.1021/nl072728k}}.
\newline\urlprefix\url{http://pubs.acs.org/cgi-bin/abstract.cgi/nalefd/2008/8/i01/abs/nl072728k.html}

\bibitem{Seibert:Nature470:78}
M.~M. Seibert, T.~Ekeberg, F.~R. N.~C. Maia, M.~Svenda, J.~Andreasson,
  O.~J{\"o}nsson, D.~Odi{\'c}, B.~Iwan, A.~Rocker, D.~Westphal, M.~Hantke,
  D.~P. Deponte, A.~Barty, J.~Schulz, L.~Gumprecht, N.~Coppola, A.~Aquila,
  M.~Liang, T.~A. White, A.~Martin, C.~Caleman, S.~Stern, C.~Abergel,
  V.~Seltzer, J.-M. Claverie, C.~Bostedt, J.~D. Bozek, S.~Boutet, A.~A.
  Miahnahri, M.~Messerschmidt, J.~Krzywinski, G.~Williams, K.~O. Hodgson, M.~J.
  Bogan, C.~Y. Hampton, R.~G. Sierra, D.~Starodub, I.~Andersson, S.~Bajt,
  M.~Barthelmess, J.~C.~H. Spence, P.~Fromme, U.~Weierstall, R.~Kirian,
  M.~Hunter, R.~B. Doak, S.~Marchesini, S.~P. Hau-Riege, M.~Frank, R.~L.
  Shoeman, L.~Lomb, S.~W. Epp, R.~Hartmann, D.~Rolles, A.~Rudenko, C.~Schmidt,
  L.~Foucar, N.~Kimmel, P.~Holl, B.~Rudek, B.~Erk, A.~H{\"o}mke, C.~Reich,
  D.~Pietschner, G.~Weidenspointner, L.~Str{\"u}der, G.~Hauser, H.~Gorke,
  J.~Ullrich, I.~Schlichting, S.~Herrmann, G.~Schaller, F.~Schopper, H.~Soltau,
  K.-U. K{\"u}hnel, R.~Andritschke, C.-D. Schr{\"o}ter, F.~Krasniqi, M.~Bott,
  S.~Schorb, D.~Rupp, M.~Adolph, T.~Gorkhover, H.~Hirsemann, G.~Potdevin,
  H.~Graafsma, B.~Nilsson, H.~N. Chapman, J.~Hajdu,
  \href{http://www.nature.com/nature/journal/v470/n7332/full/nature09748.html}{Single
  mimivirus particles intercepted and imaged with an x-ray laser}, Nature
  470~(7332) (2011) 78.
\newblock \href {http://dx.doi.org/10.1038/nature09748}
  {\path{doi:10.1038/nature09748}}.
\newline\urlprefix\url{http://www.nature.com/nature/journal/v470/n7332/full/nature09748.html}

\bibitem{Neutze:Nature406:752}
R.~Neutze, R.~Wouts, D.~van~der Spoel, E.~Weckert, J.~Hajdu,
  \href{http://dx.doi.org/10.1038/35021099}{Potential for biomolecular imaging
  with femtosecond {X}-ray pulses}, Nature 406~(6797) (2000) 752--757.
\newblock \href {http://dx.doi.org/10.1038/35021099}
  {\path{doi:10.1038/35021099}}.
\newline\urlprefix\url{http://dx.doi.org/10.1038/35021099}

\bibitem{Nass:JSR22:225}
K.~Nass, L.~Foucar, T.~R.~M. Barends, E.~Hartmann, S.~Botha, R.~L. Shoeman,
  R.~B. Doak, R.~Alonso-Mori, A.~Aquila, S.~Bajt, A.~Barty, R.~Bean, K.~R.
  Beyerlein, M.~Bublitz, N.~Drachmann, J.~Gregersen, H.~O. J{\"{o}}nsson,
  W.~Kabsch, S.~Kassemeyer, J.~E. Koglin, M.~Krumrey, D.~Mattle,
  M.~Messerschmidt, P.~Nissen, L.~Reinhard, O.~Sitsel, D.~Sokaras, G.~J.
  Williams, S.~Hau-Riege, N.~Timneanu, C.~Caleman, H.~N. Chapman, S.~Boutet,
  I.~Schlichting,
  \href{http://dx.doi.org/10.1107/S1600577515002349}{Indications of radiation
  damage in ferredoxin microcrystals using high-intensity {X-FEL} beams}, J.\
  Synchrotron\ Rad. 22~(2) (2015) 225--238.
\newblock \href {http://dx.doi.org/10.1107/S1600577515002349}
  {\path{doi:10.1107/S1600577515002349}}.
\newline\urlprefix\url{http://dx.doi.org/10.1107/S1600577515002349}

\bibitem{Lorenz:PRE86:051911}
U.~Lorenz, N.~M. Kabachnik, E.~Weckert, I.~A. Vartanyants,
  \href{http://link.aps.org/doi/10.1103/PhysRevE.86.051911}{Impact of ultrafast
  electronic damage in single-particle x-ray imaging experiments}, Phys.\ Rev.\
  E 86 (2012) 051911.
\newblock \href {http://arxiv.org/abs/1206.6960} {\path{arXiv:1206.6960}},
  \href {http://dx.doi.org/10.1103/PhysRevE.86.051911}
  {\path{doi:10.1103/PhysRevE.86.051911}}.
\newline\urlprefix\url{http://link.aps.org/doi/10.1103/PhysRevE.86.051911}

\bibitem{Ziaja:NJP14:115015}
B.~Ziaja, H.~N. Chapman, R.~F{\"a}ustlin, S.~Hau-Riege, Z.~Jurek, A.~V. Martin,
  S.~Toleikis, F.~Wang, E.~Weckert, R.~Santra,
  \href{http://stacks.iop.org/1367-2630/14/i=11/a=115015}{Limitations of
  coherent diffractive imaging of single objects due to their damage by intense
  x-ray radiation}, New J.\ Phys. 14~(11) (2012) 115015.
\newblock \href {http://dx.doi.org/10.1088/1367-2630/14/11/115015}
  {\path{doi:10.1088/1367-2630/14/11/115015}}.
\newline\urlprefix\url{http://stacks.iop.org/1367-2630/14/i=11/a=115015}

\bibitem{Bogan:AST44:i}
M.~J. Bogan, S.~Boutet, H.~N. Chapman, S.~Marchesini, A.~Barty, W.~H. Benner,
  U.~Rohner, M.~Frank, S.~P. Hau-Riege, S.~Bajt, B.~Woods, M.~M. Seibert,
  B.~Iwan, N.~Timneanu, J.~Hajdu, J.~Schulz,
  \href{http://dx.doi.org/10.1080/02786820903485800}{Aerosol imaging with a
  soft x-ray free electron laser}, Aerosol Sci. Techn. 44~(3) (2010) i--vi.
\newblock \href {http://dx.doi.org/10.1080/02786820903485800}
  {\path{doi:10.1080/02786820903485800}}.
\newline\urlprefix\url{http://dx.doi.org/10.1080/02786820903485800}

\bibitem{Ekeberg:PRL114:098102}
T.~Ekeberg, M.~Svenda, C.~Abergel, F.~R. N.~C. Maia, V.~Seltzer, J.-M.
  Claverie, M.~Hantke, O.~J{\"o}nsson, C.~Nettelblad, G.~van~der Schot,
  M.~Liang, D.~P. Deponte, A.~Barty, M.~M. Seibert, B.~Iwan, I.~Andersson,
  N.~D. Loh, A.~V. Martin, H.~Chapman, C.~Bostedt, J.~D. Bozek, K.~R. Ferguson,
  J.~Krzywinski, S.~W. Epp, D.~Rolles, A.~Rudenko, R.~Hartmann, N.~Kimmel,
  J.~Hajdu,
  \href{http://dx.doi.org/10.1103/PhysRevLett.114.098102}{Three-dimensional
  reconstruction of the giant mimivirus particle with an x-ray free-electron
  laser}, Phys.\ Rev.\ Lett. 114~(9) (2015) 098102.
\newblock \href
  {http://dx.doi.org/https://doi.org/10.1103/PhysRevLett.114.098102}
  {\path{doi:https://doi.org/10.1103/PhysRevLett.114.098102}}.
\newline\urlprefix\url{http://dx.doi.org/10.1103/PhysRevLett.114.098102}

\bibitem{Chapman:NatMater8:299}
H.~N. Chapman, \href{http://dx.doi.org/10.1038/nmat2402}{X-ray imaging beyond
  the limits}, Nature Mater. 8~(4) (2009) 299--301.
\newblock \href {http://dx.doi.org/10.1038/nmat2402}
  {\path{doi:10.1038/nmat2402}}.
\newline\urlprefix\url{http://dx.doi.org/10.1038/nmat2402}

\bibitem{DePonte:JPD41:195505}
D.~P. DePonte, U.~Weierstall, K.~Schmidt, J.~Warner, D.~Starodub, J.~C.~H.
  Spence, R.~B. Doak,
  \href{http://iopscience.iop.org/0022-3727/41/19/195505}{{Gas dynamic virtual
  nozzle for generation of microscopic droplet streams}}, J.\ Phys.\ D 41~(19)
  (2008) 195505.
\newblock \href {http://dx.doi.org/10.1088/0022-3727/41/19/195505}
  {\path{doi:10.1088/0022-3727/41/19/195505}}.
\newline\urlprefix\url{http://iopscience.iop.org/0022-3727/41/19/195505}

\bibitem{Awel:JACR51:133}
S.~Awel, R.~A. Kirian, M.~O. Wiedorn, K.~R. Beyerlein, N.~Roth, D.~A. Horke,
  D.~Oberth{\"{u}}r, J.~Knoska, V.~Mariani, A.~Morgan, L.~Adriano,
  A.~Tolstikova, P.~L. Xavier, O.~Yefanov, A.~Aquila, A.~Barty,
  S.~Roy-Chowdhury, M.~S. Hunter, D.~James, J.~S. Robinson, U.~Weierstall,
  A.~V. Rode, S.~Bajt, J.~K{\"{u}}pper, H.~N. Chapman,
  \href{https://doi.org/10.1107/S1600576717018131}{{Femtosecond X-ray
  diffraction from an aerosolized beam of protein nanocrystals}}, J.\ Appl.\
  Cryst. 51~(1) (2018) 133--139.
\newblock \href {http://arxiv.org/abs/1702.04014} {\path{arXiv:1702.04014}},
  \href {http://dx.doi.org/10.1107/S1600576717018131}
  {\path{doi:10.1107/S1600576717018131}}.
\newline\urlprefix\url{https://doi.org/10.1107/S1600576717018131}

\bibitem{Hantke:NatPhoton8:943}
M.~F. Hantke, D.~Hasse, M.~R.~N. C, T.~Ekeberg, K.~John, M.~Svenda, N.~D. Loh,
  A.~V. Martin, N.~Timneanu, L.~S. D, van~der SchotGijs, G.~H. Carlsson,
  M.~Ingelman, J.~Andreasson, D.~Westphal, M.~Liang, F.~Stellato, D.~P.
  DePonte, R.~Hartmann, N.~Kimmel, R.~A. Kirian, M.~M. Seibert, K.~M{\"u}hlig,
  S.~Schorb, K.~Ferguson, C.~Bostedt, S.~Carron, J.~D. Bozek, D.~Rolles,
  A.~Rudenko, S.~Epp, H.~N. Chapman, A.~Barty, J.~Hajdu, I.~Andersson,
  \href{http://www.nature.com/nphoton/journal/v8/n12/full/nphoton.2014.270.html}{High-throughput
  imaging of heterogeneous cell organelles with an x-ray laser}, Nat. Photon.
  8~(12) (2014) 943--949.
\newblock \href {http://dx.doi.org/doi:10.1038/nphoton.2014.270}
  {\path{doi:doi:10.1038/nphoton.2014.270}}.
\newline\urlprefix\url{http://www.nature.com/nphoton/journal/v8/n12/full/nphoton.2014.270.html}

\bibitem{Kirian:SD2:041717}
R.~A. Kirian, S.~Awel, N.~Eckerskorn, H.~Fleckenstein, M.~Wiedorn, L.~Adriano,
  S.~Bajt, M.~Barthelmess, R.~Bean, K.~R. Beyerlein, L.~M.~G. Chavas,
  M.~Domaracky, M.~Heymann, D.~A. Horke, J.~Knoska, M.~Metz, A.~Morgan,
  D.~Oberthuer, N.~Roth, T.~Sato, P.~L. Xavier, O.~Yefanov, A.~V. Rode,
  J.~K{\"u}pper, H.~N. Chapman,
  \href{http://dx.doi.org/10.1063/1.4922648}{Simple convergent-nozzle aerosol
  injector for single-particle diffractive imaging with x-ray free-electron
  lasers}, Struct.\ Dyn. 2~(4) (2015) 041717.
\newblock \href {http://dx.doi.org/10.1063/1.4922648}
  {\path{doi:10.1063/1.4922648}}.
\newline\urlprefix\url{http://dx.doi.org/10.1063/1.4922648}

\bibitem{Awel:OE24:6507}
S.~Awel, R.~A. Kirian, N.~Eckerskorn, M.~Wiedorn, D.~A. Horke, A.~V. Rode,
  J.~K\"upper, H.~N. Chapman,
  \href{http://dx.doi.org/10.1364/OE.24.006507}{{V}isualizing aerosol-particle
  injection for diffractive-imaging experiments}, Opt.\ Exp. 24~(6) (2016)
  6507--6521.
\newblock \href {http://dx.doi.org/10.1364/OE.24.006507}
  {\path{doi:10.1364/OE.24.006507}}.
\newline\urlprefix\url{http://dx.doi.org/10.1364/OE.24.006507}

\bibitem{Bergh:QRB41:181}
M.~Bergh, G.~Huldt, N.~Tîmneanu, F.~R. N.~C. Maia, J.~Hajdu,
  \href{http://journals.cambridge.org/article_S003358350800471X}{Feasibility of
  imaging living cells at subnanometer resolutions by ultrafast x-ray
  diffraction}, Quarterly Reviews of Biophysics 41 (2008) 181--204.
\newblock \href {http://dx.doi.org/10.1017/S003358350800471X}
  {\path{doi:10.1017/S003358350800471X}}.
\newline\urlprefix\url{http://journals.cambridge.org/article_S003358350800471X}

\bibitem{Fung:Nature532:471}
R.~Fung, A.~M. Hanna, O.~Vendrell, S.~Ramakrishna, T.~Seideman, R.~Santra,
  A.~Ourmazd, \href{http://dx.doi.org/10.1038/nature17627}{{Dynamics from noisy
  data with extreme timing uncertainty}}, Nature 532~(7600) (2016) 471--475.
\newblock \href {http://dx.doi.org/10.1038/nature17627}
  {\path{doi:10.1038/nature17627}}.
\newline\urlprefix\url{http://dx.doi.org/10.1038/nature17627}

\bibitem{Barty:ARPC64:415}
A.~Barty, J.~K{\"u}pper, H.~N. Chapman,
  \href{http://dx.doi.org/10.1146/annurev-physchem-032511-143708}{Molecular
  imaging using x-ray free-electron lasers}, Annu.\ Rev.\ Phys.\ Chem. 64~(1)
  (2013) 415--435.
\newblock \href {http://dx.doi.org/10.1146/annurev-physchem-032511-143708}
  {\path{doi:10.1146/annurev-physchem-032511-143708}}.
\newline\urlprefix\url{http://dx.doi.org/10.1146/annurev-physchem-032511-143708}

\bibitem{Robinson:CPAM9:69}
A.~Robinson,
  \href{http://onlinelibrary.wiley.com/doi/10.1002/cpa.3160090105/abstract}{On
  the motion of small particles in a potential field of flow}, Comm.\ Pure\
  Appl.\ Math. 9~(1) (1956) 69--84.
\newblock \href {http://dx.doi.org/10.1002/cpa.3160090105}
  {\path{doi:10.1002/cpa.3160090105}}.
\newline\urlprefix\url{http://onlinelibrary.wiley.com/doi/10.1002/cpa.3160090105/abstract}

\bibitem{Liu:AST22:293}
P.~Liu, P.~J. Ziemann, D.~B. Kittelson, P.~H. McMurry,
  \href{http://dx.doi.org/10.1080/02786829408959748}{Generating particle beams
  of controlled dimensions and divergence: I. theory of particle motion in
  aerodynamic lenses and nozzle expansions}, Aerosol Sci. Techn. 22~(3) (1995)
  293--313.
\newblock \href {http://dx.doi.org/10.1080/02786829408959748}
  {\path{doi:10.1080/02786829408959748}}.
\newline\urlprefix\url{http://dx.doi.org/10.1080/02786829408959748}

\bibitem{Zhang:AST36:617}
X.~F. Zhang, K.~A. Smith, D.~R. Worsnop, J.~Jimenez, J.~T. Jayne, C.~E. Kolb,
  \href{http://www.tandfonline.com/doi/abs/10.1080/02786820252883856}{{A
  numerical characterization of particle beam collimation by an aerodynamic
  lens-nozzle system: Part I. An individual lens or nozzle}}, Aerosol Sci.
  Techn. 36~(5) (2002) 617--631.
\newblock \href {http://dx.doi.org/10.1080/02786820252883856}
  {\path{doi:10.1080/02786820252883856}}.
\newline\urlprefix\url{http://www.tandfonline.com/doi/abs/10.1080/02786820252883856}

\bibitem{Zhang:AST38:619}
X.~Zhang, K.~A. Smith, D.~R. Worsnop, J.~L. Jimenez, J.~T. Jayne, C.~E. Kolb,
  J.~Morris, P.~Davidovits,
  \href{http://www.tandfonline.com/doi/abs/10.1080/02786820490479833}{{Numerical
  Characterization of Particle Beam Collimation: Part II Integrated
  Aerodynamic-Lens{\textendash}Nozzle System}}, Aerosol Sci. Techn. 38~(6)
  (2004) 619--638.
\newblock \href {http://dx.doi.org/10.1080/02786820490479833}
  {\path{doi:10.1080/02786820490479833}}.
\newline\urlprefix\url{http://www.tandfonline.com/doi/abs/10.1080/02786820490479833}

\bibitem{Wang:AST39:624}
X.~Wang, A.~Gidwani, S.~L. Girshick, P.~H. McMurry,
  \href{http://www.tandfonline.com/doi/abs/10.1080/02786820500181950}{{Aerodynamic
  Focusing of Nanoparticles: II. Numerical Simulation of Particle Motion
  Through Aerodynamic Lenses}}, Aerosol Sci. Techn. 39~(7) (2005) 624--636.
\newblock \href {http://dx.doi.org/10.1080/02786820500181950}
  {\path{doi:10.1080/02786820500181950}}.
\newline\urlprefix\url{http://www.tandfonline.com/doi/abs/10.1080/02786820500181950}

\bibitem{Wang:AST40:320}
X.~Wang, P.~H. McMurry, \href{http://dx.doi.org/10.1080/02786820600615063}{A
  design tool for aerodynamic lens systems}, Aerosol Sci. Technol. 40~(5)
  (2006) 320--334.
\newblock \href {http://dx.doi.org/10.1080/02786820600615063}
  {\path{doi:10.1080/02786820600615063}}.
\newline\urlprefix\url{http://dx.doi.org/10.1080/02786820600615063}

\bibitem{Lee:JAS39:287}
K.-S. Lee, S.-W. Cho, D.~Lee,
  \href{http://dx.doi.org/10.1016/j.jaerosci.2007.10.011}{Development and
  experimental evaluation of aerodynamic lens as an aerosol inlet of single
  particle mass spectrometry}, J.\ Aerosol\ Sci. 39~(4) (2008) 287--304.
\newblock \href {http://dx.doi.org/10.1016/j.jaerosci.2007.10.011}
  {\path{doi:10.1016/j.jaerosci.2007.10.011}}.
\newline\urlprefix\url{http://dx.doi.org/10.1016/j.jaerosci.2007.10.011}

\bibitem{Meinen:AST44:316}
J.~Meinen, S.~Khasminskaya, E.~R{\"u}hl, W.~Baumann, T.~Leisner,
  \href{http://www.tandfonline.com/doi/abs/10.1080/02786821003639692}{The
  {TRAPS} apparatus: {E}nhancing target density of nanoparticle beams in vacuum
  for x-ray and optical spectroscopy}, Aerosol Sci. Techn. 44~(4) (2010)
  316--328.
\newblock \href {http://dx.doi.org/10.1080/02786821003639692}
  {\path{doi:10.1080/02786821003639692}}.
\newline\urlprefix\url{http://www.tandfonline.com/doi/abs/10.1080/02786821003639692}

\bibitem{Zherebtsov:NatPhys7:656}
S.~Zherebtsov, T.~Fennel, J.~Plenge, E.~Antonsson, I.~Znakovskaya, A.~Wirth,
  O.~Herrwerth, F.~S{\"u}{\ss}mann, C.~Peltz, I.~Ahmad, S.~A. Trushin,
  V.~Pervak, S.~Karsch, M.~J.~J. Vrakking, B.~Langer, C.~Graf, M.~I. Stockman,
  F.~Krausz, E.~R{\"u}hl, M.~F. Kling,
  \href{http://dx.doi.org/10.1038/nphys1983}{Controlled near-field enhanced
  electron acceleration from dielectric nanospheres with intense few-cycle
  laser fields}, Nat. Phys. 7~(8) (2011) 656--662.
\newblock \href {http://dx.doi.org/10.1038/nphys1983}
  {\path{doi:10.1038/nphys1983}}.
\newline\urlprefix\url{http://dx.doi.org/10.1038/nphys1983}

\bibitem{Canagaratna:MSR26:185}
M.~R. Canagaratna, J.~T. Jayne, J.~L. Jimenez, J.~D. Allan, M.~R. Alfarra,
  Q.~Zhang, T.~B. Onasch, F.~Drewnick, H.~Coe, A.~Middlebrook, A.~Delia, L.~R.
  Williams, A.~M. Trimborn, M.~J. Northway, P.~F. DeCarlo, C.~E. Kolb,
  P.~Davidovits, D.~R. Worsnop,
  \href{http://dx.doi.org/10.1002/mas.20115}{Chemical and microphysical
  characterization of ambient aerosols with the aerodyne aerosol mass
  spectrometer}, Mass\ Spectrom.\ Rev. 26~(2) (2007) 185--222.
\newblock \href {http://dx.doi.org/10.1002/mas.20115}
  {\path{doi:10.1002/mas.20115}}.
\newline\urlprefix\url{http://dx.doi.org/10.1002/mas.20115}

\bibitem{Comsol:Multiphysics:5.3}
COMSOL Multiphysics v.\ 5.3.\ \url{http://www.comsol.com}. COMSOL AB,
  Stockholm, Sweden.

\bibitem{Hutchins:AST22:202}
D.~K. Hutchins, M.~H. Harper, R.~L. Felder,
  \href{http://www.tandfonline.com/doi/abs/10.1080/02786829408959741}{Slip
  correction measurements for solid spherical particles by modulated dynamic
  light scattering}, Aerosol Sci. Techn. 22~(2) (1995) 202--218.
\newblock \href {http://dx.doi.org/10.1080/02786829408959741}
  {\path{doi:10.1080/02786829408959741}}.
\newline\urlprefix\url{http://www.tandfonline.com/doi/abs/10.1080/02786829408959741}

\bibitem{Li:AST16:209}
A.~Li, G.~Ahmadi,
  \href{http://www.tandfonline.com/doi/abs/10.1080/02786829208959550}{Dispersion
  and deposition of spherical particles from point sources in a turbulent
  channel flow}, Aerosol Sci. Techn. 16~(24) (1992) 209--226.
\newblock \href {http://dx.doi.org/10.1080/02786829208959550}
  {\path{doi:10.1080/02786829208959550}}.
\newline\urlprefix\url{http://www.tandfonline.com/doi/abs/10.1080/02786829208959550}

\bibitem{Beyerlein:RSI86:125104}
K.~R. Beyerlein, L.~Adriano, M.~Heymann, R.~Kirian, J.~Knoska, F.~Wilde, H.~N.
  Chapman, S.~Bajt, \href{http://dx.doi.org/10.1063/1.4936843}{Ceramic
  micro-injection molded nozzles for serial femtosecond crystallography sample
  delivery}, Rev.\ Sci.\ Instrum. 86~(12) (2015) 125104--12.
\newblock \href {http://dx.doi.org/10.1063/1.4936843}
  {\path{doi:10.1063/1.4936843}}.
\newline\urlprefix\url{http://dx.doi.org/10.1063/1.4936843}

\bibitem{Bielecki:privcomm:2017}
J.~Bielecki, private communication (2017).

\bibitem{Giuseppe:AST33:105}
G.~A. Petrucci, P.~B. Farnsworth, P.~Cavalli, N.~Omenetto,
  \href{http://dx.doi.org/10.1080/027868200410877}{A differentially pumped
  particle inlet for sampling of atmospheric aerosols into a time-of-flight
  mass spectrometer: Optical characterization of the particle beam}, Aerosol
  Sci. Techn. 33~(1-2) (2000) 105--121.
\newblock \href {http://dx.doi.org/10.1080/027868200410877}
  {\path{doi:10.1080/027868200410877}}.
\newline\urlprefix\url{http://dx.doi.org/10.1080/027868200410877}

\bibitem{Headrick:JAS58:158}
J.~M. Headrick, P.~E. Schrader, H.~A. Michelsen,
  \href{http://www.sciencedirect.com/science/article/pii/S0021850213000062}{Radial-profile
  and divergence measurements of combustion-generated soot focused by an
  aerodynamic-lens system}, J.\ Aerosol\ Sci. 58 (2013) 158--170.
\newblock \href
  {http://dx.doi.org/http://dx.doi.org/10.1016/j.jaerosci.2013.01.002}
  {\path{doi:http://dx.doi.org/10.1016/j.jaerosci.2013.01.002}}.
\newline\urlprefix\url{http://www.sciencedirect.com/science/article/pii/S0021850213000062}

\bibitem{Horke:JAP121:123106}
D.~A. Horke, N.~Roth, L.~Worbs, J.~Küpper,
  \href{http://dx.doi.org/10.1063/1.4978914}{Characterizing gas flow from
  aerosol particle injectors}, J.\ Appl.\ Phys. 121~(12) (2017) 123106.
\newblock \href {http://arxiv.org/abs/1609.09020} {\path{arXiv:1609.09020}},
  \href {http://dx.doi.org/10.1063/1.4978914} {\path{doi:10.1063/1.4978914}}.
\newline\urlprefix\url{http://dx.doi.org/10.1063/1.4978914}

\bibitem{Filsinger:PRL100:133003}
F.~Filsinger, U.~Erlekam, G.~von Helden, J.~K{\"u}pper, G.~Meijer,
  \href{http://dx.doi.org/10.1103/PhysRevLett.100.133003}{Selector for
  structural isomers of neutral molecules}, Phys.\ Rev.\ Lett. 100 (2008)
  133003.
\newblock \href {http://arxiv.org/abs/0802.2795} {\path{arXiv:0802.2795}},
  \href {http://dx.doi.org/10.1103/PhysRevLett.100.133003}
  {\path{doi:10.1103/PhysRevLett.100.133003}}.
\newline\urlprefix\url{http://dx.doi.org/10.1103/PhysRevLett.100.133003}

\bibitem{Chang:IRPC34:557}
Y.-P. Chang, D.~A. Horke, S.~Trippel, J.~K{\"u}pper,
  \href{http://dx.doi.org/10.1080/0144235X.2015.1077838}{Spatially-controlled
  complex molecules and their applications}, Int.\ Rev.\ Phys.\ Chem. 34 (2015)
  557--590.
\newblock \href {http://arxiv.org/abs/1505.05632} {\path{arXiv:1505.05632}},
  \href {http://dx.doi.org/10.1080/0144235X.2015.1077838}
  {\path{doi:10.1080/0144235X.2015.1077838}}.
\newline\urlprefix\url{http://dx.doi.org/10.1080/0144235X.2015.1077838}

\end{thebibliography}
\bibliographystyle{elsarticle-num}
\end{document}